\documentclass[onecolumn,aps,nofootinbib,superscriptaddress]{revtex4}
\usepackage[colorlinks, linkcolor=blue, anchorcolor=blue, citecolor=blue, colorlinks=true, urlcolor=blue]{hyperref}
\usepackage{amssymb}
\usepackage{amsmath}
\usepackage{graphicx}
\usepackage{lscape}
\usepackage{booktabs}
\usepackage{multirow}
\begin{document}
\title{Study for (anti)hypertriton and light (anti)nuclei productions in high energy collisions at $\sqrt{S_{NN}}$ = 200 GeV}

\author{Hai-Jun Li}
\email{lhj718@foxmail.com}
\affiliation{School of Mathematics and Physics, China University of Geosciences, Wuhan 430074, China}
\author{Ting-Ting Zeng}
\affiliation{School of Mathematics and Physics, China University of Geosciences, Wuhan 430074, China}
\author{Gang Chen}
\affiliation{School of Mathematics and Physics, China University of Geosciences, Wuhan 430074, China}

\begin{abstract}
We use the parton and hadron cascade (PACIAE) model and the dynamically constrained phase-space coalescence (DCPC) model to investigate the productions of (anti)hypertriton and light (anti)nuclei generated by 0-10\% centrality $^{12}$C+$^{12}$C, $^{24}$Mg+$^{24}$Mg, $^{40}$Ca+$^{40}$Ca, and $^{64}$Cu+$^{64}$Cu collisions at $\sqrt{S_{NN}}$ = 200 GeV with $|y|$ $<$ 1.5 and $p_{T}$ $<$ 5. 
We study the yield ratios of the antiparticle to particle and the rapidity distributions of the different (anti)nuclei. 
We find that the amounts of antimatter produced are significantly lower than that of the corresponding particles, the results of theoretical models are well consistent with the PHOBOS data. 
The yield ratios of the particle to antiparticle in different transverse momentum regions are also given, and we find that the ratios are raise with the increase of the transverse momentum.

\end{abstract}

\maketitle

\section{Introduction}
In 1928, Dirac predicted the existence of negative energy states (i.e., antimatter) of the electrons depending on the symmetry principle of the quantum mechanics for the first time~\cite{pamd}. 
According to the Big Bang theory~\cite{1}, it is generally accepted that equal amounts of matter and antimatter have been produced during the initial state of the Universe. 
However, this symmetry got lost in a series of evolutions of the Universe with no significant amount of antimatter exist now~\cite{2}. 
Since the high energy ultrarelativistic heavy-ion collisions could create abundant hyperons and nucleons, and the initial fireballs produced by the heavy-ion collisions are very similar to the Universe initial state, so the study of antinuclei productions in heavy-ion collisions maybe a better choice to solve above puzzle~\cite{3}. 
However, since the low abundance of the light nuclei (antinuclei) and the antimatter are unstable, the research of their productions may be quite difficult both in theoretically and experimentally~\cite{4}.

The antimatter nuclei have been widely studied in cosmic ray~\cite{5,6} and accelerator experiments~\cite{7,8}, which could be the indirect signals of the new physics, such as the dark matter (DM) and the manmade matter called quark gluon plasma (QGP)~\cite{9,10}.
Hypernuclei provide a doorway to the productions of strangelets and a unique opportunity to study the hyperon-nucleon (YN) and hyperon-hyperon (YY) interactions~\cite{11}. 
The measurements of YN and YY interactions are essential for the theoretical study of neutron stars~\cite{12} and exotic states of finite nuclei~\cite{13}. 
In 2010, the STAR Collaboration reported their measurements of $\rm {_{\Lambda}^3H}$ and $\rm {\overline{_{\overline\Lambda}^3H}}$ productions in Au+Au collisions at top Relativistic Heavy-Ion Collider (RHIC) energy~\cite{14}, including 70$\pm$17 antihypertritons ($\rm {\overline{_{\overline\Lambda}^3H}}$) and 157$\pm$30 hypertritons ($\rm {_{\Lambda}^3H}$) in the 8.9$\times$10$^7$ minimum-bias and 2.2$\times$10$^7$ central (head-on) Au+Au collision events. 
Additionally, they also reported 18 $\rm ^4He$ counts at RHIC in $10^9$ recorded Au+Au collisions at the center-of-mass energies of 200 GeV and 62 GeV per nucleon-nucleon pair in 2011~\cite{15}. 
Therefore, we have the simulation study about the productions of (anti)hypertriton and light (anti)nuclei at high energy (200 GeV) in $^{12}$C+$^{12}$C, $^{24}$Mg+$^{24}$Mg, $^{40}$Ca+$^{40}$Ca, and $^{64}$Cu+$^{64}$Cu collisions.

We have proposed a dynamically constrained phase-space coalescence (DCPC) model~\cite{16} based on the parton and hadron cascade (PACIAE) model~\cite{17,18}. 
We have predicted the light nuclei (antinuclei) yields, transverse momentum distributions, and the rapidity distributions in $pp$ collisions at 7 and 14 TeV~\cite{16}. 
We also have investigated the light nuclei (antinuclei), hypernuclei (antihypernuclei) productions~\cite{4} and studied their centrality dependence~\cite{19}, scaling feature~\cite{20} in the 0-5\% most central Au+Au collisions at 200 GeV. 
Thus, these works make the study for the (anti)hypertriton and light (anti)nuclei productions in nucleon-nucleon (NN) collisions possibile.

This paper is structured as follows. In Sec.~\ref{model}, the PACIAE model coupled with the DCPC model are developed to study the production rates of hypertriton ($\rm {_{\Lambda}^3H}$) and light nuclei, i.e. $d$, $t$, and
$\rm ^3He$  corresponding to their antimatter in high energy collisions.
In Sec.~\ref{res}, we calculate the yield ratios of antiparticle to particle in $^{12}$C+$^{12}$C, $^{24}$Mg+$^{24}$Mg, $^{40}$Ca+$^{40}$Ca, and $^{64}$Cu+$^{64}$Cu collisions with the beams of energy at $\sqrt{S_{NN}}$ $=$ 200 GeV.
The studies of the rapidity distributions and the transverse momentum distributions of $\bar{d}/d$, $\bar{t}/t$, $\rm {\overline{^3He}}/^3He$ and $\rm {\overline{_{\overline\Lambda}^3H}}/{_{\overline\Lambda}^3H}$ are also given. 
The conclusion is given in Sec.~\ref{sum}.

\section {MODELS}
\label{model}

In this section, we briefly introduce the PACIAE and DCPC models.
The PYTHIA model (PYTHIA 6.4~\cite{21}) is designed for the high energy hadron-hadron collisions. 
The process of hadron-hadron collisions can be described by the parton-parton collisions in PYTHIA. 
The QCD radiation initial/final states and the multi-parton interactions are also considered. 
The final hadron states can be eventually derived by the hadron-hadron collisions.
The PACIAE~\cite{17,18} model is based on PYTHIA 6.4, which is designed for the nucleus-nucleus collisions. 
In PACIAE, the nucleus-nucleus collisions can be described by the nucleon-nucleon (NN) collisions. 
The initial parton state, composed of quarks/antiquarks and gluons, is regarded as the quark-gluon matter (QGM) generated in the relativistic nucleus-nucleus collision. 
The rescattering among the QGM partons can be described by the $2\rightarrow 2$ Lo-pQCD parton-parton cross sections~\cite{22}. 
In order to describe the high order and the nonperturbative corrections, the $K$ factor is added in the model. 
The hadronic matter continues to scatter until hadronic freeze-out~\cite{23,24}.
More details about these models can be found in Ref.~\cite{18}.

According to the uncertainty principle of the quantum statistical mechanics
\begin{equation}
\Delta\vec q\Delta\vec p\geqslant h^3, 
\end{equation}
we could not give the precisely informations of both the position $\vec q= (x,y,z)$ and the momentum $\vec p = (p_x,p_y,p_z)$ of a particle in the six-dimension phase space.
The yield of a single particle can be described by
\begin{equation}
\int_{\mathcal{H}\leqslant \mathcal{E}} \frac{d\vec qd\vec p}{h^3},
\end{equation}
with the Hamiltonian $\mathcal{H}$ and the energy $\mathcal{E}$ of the particle.
Thus, we can only derive the particle lies inside a quantum ``box" or ``state" in a six-dimension phase space with a phase space volume of $\Delta\vec q\Delta\vec p$.
Additionally, we can derive the yield for the $N$-particle system
\begin{equation}
\int ...\int_{\mathcal{H}\leqslant \mathcal{E}} \frac{d\vec q_1d\vec p_1...d\vec
q_Nd\vec p_N}{h^{3N}}.
\label{phas}
\end{equation}

In this work, in order to obtain the yield of $\rm \overline{_{\overline\Lambda}^3H}$ in the DCPC model, we have
\begin{align}
y=&\int ... \int\delta_{123}\frac{d\vec q_1d\vec p_1
  d\vec q_2d\vec p_2d\vec q_3d\vec p_3}{h^{9}},
\label{yield} \\
\delta_{123}=&\left\{
  \begin{array}{ll}
  1 \hspace{0.4cm} \textrm{if} \hspace{0.2cm} 1= \bar p,\hspace{0.2cm} 2= \bar n,\hspace{0.2cm}
    3= \bar\Lambda,\hspace{0.2cm}  {\rm combination};\\
    \hspace{0.6cm} m_0\leqslant m_{\rm inv}\leqslant m_0+\Delta m;\\
    \hspace{0.6cm} |q_{ij}|\leqslant D_0,(i\neq j;i,j=1,2,3); \\
  0 \hspace{0.4cm}\textrm{otherwise};
  \end{array}
  \right.
\label{funct5}
\end{align}
with the rest mass of $\rm \overline{_{\overline\Lambda}^3H}$, $m_0$, the diameter $D_0$, the allowed mass uncertainty $\Delta m$, and the vector distance between particles $i$ and $j$, $|\vec q_{ij}|=|\vec q_{i}-\vec q_{j}|$, where 
\begin{equation}
m_{\rm inv}=\Bigg[\bigg(\sum^{3}_{i=1} \mathcal{E}_i \bigg)^2-\bigg(\sum^{3}_{i=1}
\vec p_i \bigg)^2 \Bigg]^{1/2},
\end{equation}
with the particle energy ($\mathcal{E}_{1}$, $\mathcal{E}_{2}$, $\mathcal{E}_{3}$) and momentum ($\vec p_1$, $\vec p_2$, $\vec p_3$).
Then we can derive the configuration of the $\rm {\overline{_{\overline\Lambda}^3H}}$ ($\bar p$$+$$\bar n$$+$$\bar\Lambda$) system from a single event
\begin{equation}
C_{\bar p\bar n\bar\Lambda}(\Delta q_i; \hspace{0.1cm}\vec p_i),(i=1,2,3),
\label{funct7}
\end{equation}
The diameter constraint in Eq.~(\ref{funct5}) can be taken as
\begin{equation}
\Delta q_i\leqslant R_0, (i=1,2,3),
\end{equation}
with the radius of $\rm \overline{_{\overline\Lambda}^3H}$, $R_0$.

Above configurations contribute the partial yield to $\rm \overline{_{\overline\Lambda}^3H}$ with
\begin{equation}
y_{123}=\left\{
  \begin{array}{ll}
  1 \hspace{0.4cm} \textrm{if} \hspace{0.2cm} m_0\leqslant m_{\rm inv}\leqslant m_0+\Delta m,\\
  0 \hspace{0.4cm}\textrm{otherwise};
  \end{array}
  \right.
\label{yield3}
\end{equation}
Therefore, we can derive the whole yield of $\rm \overline{_{\overline\Lambda}^3H}$ from a single event by the sum of the partial yield  configurations of Eq.~(\ref{funct7}) and the combinations.

\begin{table}[tbp]
\caption{The integrated yield dN/dy of particles with $0-10\%$ centrality, for $p$, $\bar{p}$, $\Lambda$, and $\bar{\Lambda}$ in $^{64}$Cu+$^{64}$Cu collisions at $\sqrt{S_{NN}}$ $=$ 62.4 GeV and $\sqrt{S_{NN}}$ $=$ 200 GeV.}
\setlength{\tabcolsep}{ 29.5pt}
\renewcommand{\arraystretch}{2}
\begin{tabular}{c|cc|cc} \hline \hline
\multirow{2}{*}{Particle type}& \multicolumn{2}{c|}{62.4(GeV)}&\multicolumn{2}{c}{200(GeV)} \\
\cline{2-3} \cline{4-5} &  PACIAE &  PHENIX$^a$ &  PACIAE & STAR$^b$ \\ 
\hline 
$p$  & 4.49 &  $4.54$ & $7.48$ &$ $ \\ $\Bar{ p}$ & $2.57$& $2.73$ & $4.82 $ &  \\ 
$\Lambda$  &$2.18$& $ $ & $4.75$ &$4.68\pm0.45$\\ 
$\Bar{\Lambda }$   &$1.45$& $ $ & $3.58$ &$3.79\pm0.37$ \\ 
\hline 
\hline
\multicolumn{4}{l}{$^a$ The PHENIX data are taken from Ref.~\cite{27}.} \\
\multicolumn{4}{l}{$^b$ The STAR data are taken from Ref.~\cite{28}.} \\
\end{tabular} \label{biao1}
\end{table}

\section {Results and Discussion}
\label{res}

We use the PACIAE~\cite{17,18} model to generate the final state of the particles.
The hyperons are heavier than $\Lambda$ in the PACIAE model.
The values of the model parameters are fixed with the values given by PYTHIA model. 
Additionally, the values of the parameters $K$ factor, parj(1), parj(2), and parj(3), corresponding to the PYTHIA strange productions, are given by fitting the PHENIX data of $p$, $\bar{p}$ in Cu+Cu collisions at $\sqrt{S_{NN}}$ $=$ 62.4 GeV by the 0-10\% centrality with 3.0 $<$ $|y|$ $<$ 3.9 and 0.5 $<$ $p_{T}$ $<$ 4.5~\cite{27} and the STAR data of $\Lambda$, $\bar{\Lambda}$ in Cu+Cu collisions at $\sqrt{S_{NN}}$ $=$ 200 GeV by the $0-10\%$ centrality with $|y|$ $<$ 0.5 and $p_{T}$ $<$ 5~\cite{28}, which are shown in Tab.~\ref{biao1}. 
Then we have the fitted parameters with $K$ = 3, parj(1) = 0.15, parj(2) = 0.45, and parj(3) = 0.65. 
The yields dN/dy of $\Bar{d}$ is calculated by the DCPC model in Cu+Cu collisions at $\sqrt{S_{NN}}$ $=$ 200 GeV by $0-10\%$ centrality with $|y|$ $<$ 0.9 and 0.2 $<$ $p_{T}$ $<$ 0.9, we calculate the $\Delta m$ = 0.0003.
The results from our model agree well with the experimental data from the STAR Collaboration \cite{28}.

We generate $5\times10^7$ minimum-bias events by the PACIAE model in $^{12}$C+$^{12}$C, $^{24}$Mg+$^{24}$Mg, $^{40}$Ca+$^{40}$Ca, and $^{64}$Cu+$^{64}$Cu collisions at $\sqrt{S_{NN}}$ $=$ 200 GeV, the
rapidity region we selected is $|y|$ $<$ 1.5, and the transverse momentum region is 0 $<$ $p_{T}$ $<$ 5 (GeV/c). 
The integrated yields dN/dy of (anti)hypertriton $\rm {_{\Lambda}^3 H}$ ($\rm \overline{_{\overline\Lambda}^3H}$) and light (anti)nuclei $d$ ($\overline d$), $t$ ($\overline t$), and $\rm ^3{He}$ ($\rm {\overline{^3He}}$) are simulated by the DCPC model, which are shown in Tab.~\ref{biao2}.

\begin{figure}[htbp]
\includegraphics[width=0.75\textwidth]{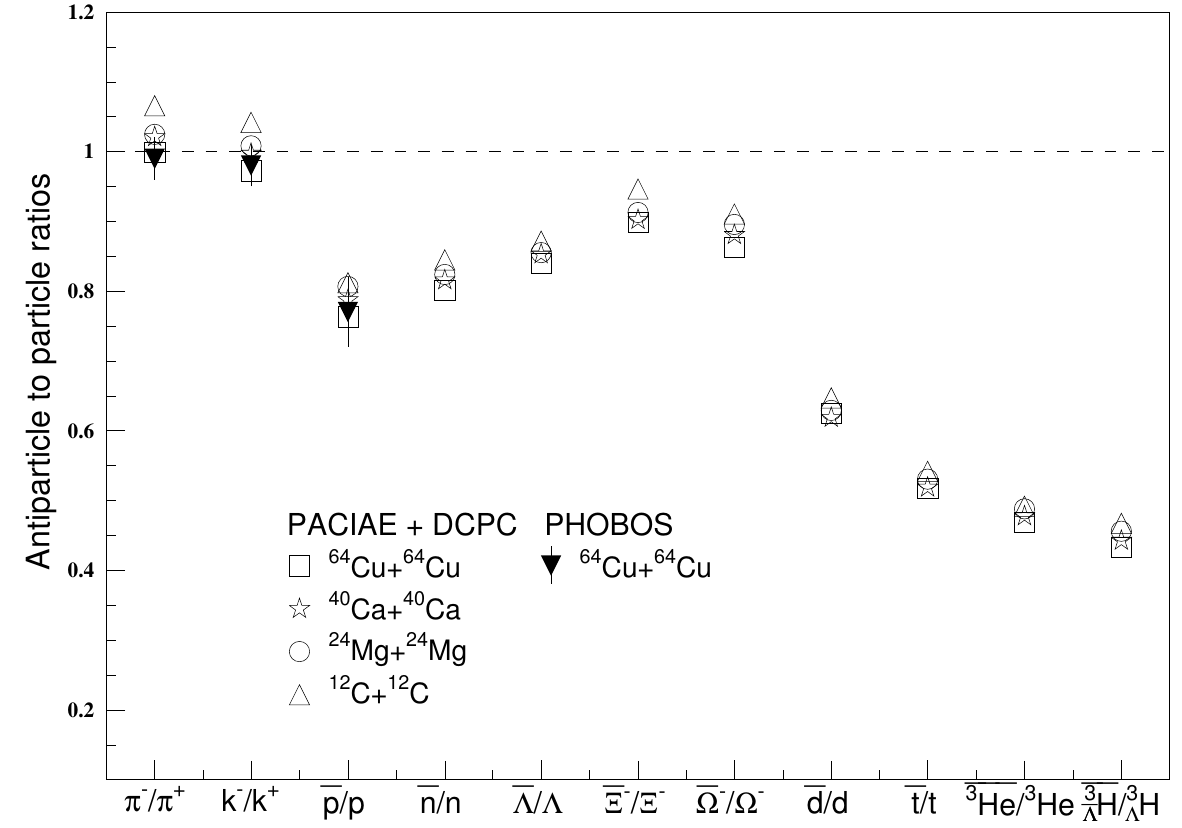}
\caption{The yield ratios of particles ($\pi^+$, $K^+$, $p$, $n$,
$\Lambda$, $\Xi^-$, and $\Omega^-$) to corresponding antiparticles ($\pi^-$, $K^-$, $\bar{p}$, $\bar{n}$, $\bar{\Lambda}$, ${\overline{\Xi^-}}$, and ${\overline{\Omega^-}}$) in $0-10\%$ centrality $^{12}$C+$^{12}$C, $^{24}$Mg+$^{24}$Mg, $^{40}$Ca+$^{40}$Ca, and $^{64}$Cu+$^{64}$Cu at $\sqrt{S_{NN}}$ $=$ 200 GeV reaction. Solid symbols are the experimental data points from the PHOBOS Collaboration \cite{30}. 
Open symbols represent the PACIAE + DCPC model results.} 
\label{tu1}
\end{figure}

The yield ratios of antiparticle to particle are given in FIG.~\ref{tu1}, including $\pi^-/\pi^+$, $K^-/K^+$, $\bar{p}/p$, $\bar{n}/n$, $\bar{\Lambda}/\Lambda$, ${\overline{\Xi^-}}/\Xi^-$, and ${\overline{\Omega^-}}/\Omega^-$. 
For the different collision nucleon, $^{12}$C+$^{12}$C, $^{24}$Mg+$^{24}$Mg, $^{40}$Ca+$^{40}$Ca, and $^{64}$Cu+$^{64}$Cu, the ratios of antiparticle to particle are roughly unchanged. 
We find that the bigger nucleon number of the collisions nucleon corresponding to the lower yield ratios of antiparticles to particles. 
It means that the productions of antimatter in the collisions of light nuclei are easier in the same conditions, and this is suit for the nucleus and particles productions.
We can also see that the particles ($\pi^+$, $K^+$, $p$, $n$, $\Lambda$, $\Xi^-$, and $\Omega^-$) ratios approach to 1, while the hypertriton ($\rm {_{\Lambda}^3H}$) and light nuclei ($d$, $t$, and $\rm ^3He$) is less than 1. The results obtained from our models are in agreement with the experimental data from the PHOBOS Collaboration \cite{30}.

\begin{table}[t]
\caption{The integrated yield dN/dy of particles at midrapidity ($|y|$ $<$ 0.5) with $0-10\%$ centrality, for $d$ ($\overline d$), $t$ ($\overline t$), $\rm ^3{He}$ ($\rm {\overline{^3He}}$) and $\rm {_{\Lambda}^3 H}$ ($\rm \overline{_{\overline\Lambda}^3H}$) in $^{12}$C+$^{12}$C, $^{24}$Mg+$^{24}$Mg, $^{40}$Ca+$^{40}$Ca, and $^{64}$Cu+$^{64}$Cu collisions at $\sqrt{S_{NN}}$ $=$ 200 GeV.}
\setlength{\tabcolsep}{ 7.1pt}
\renewcommand{\arraystretch}{2}
\begin{tabular}{ccccccccc} \hline \hline
Collision types& $d$ & $\bar{d}$ & $t$  &$\bar{t}$ &$\rm ^3{He}$ &$\rm {\overline{^3He}}$& $\rm {_{\Lambda}^3 H}$ &$\rm \overline{_{\overline\Lambda}^3H}$ \\
\hline
$^{12}$C+$^{12}$C&   2.187E-03 & 1.415E-03 & 4.227E-07  & 2.292E-07 & 3.492E-07 & 1.721E-07 & 2.364E-07 & 1.104E-07 \\
$^{24}$Mg+$^{24}$Mg& 7.301E-03 & 4.598E-03 & 1.114E-06  & 5.907E-07 & 9.121E-07 & 4.455E-07 & 6.453E-07 & 2.942E-07 \\
$^{40}$Ca+$^{40}$Ca& 2.094E-02 & 1.298E-02 & 2.676E-06  & 1.391E-06 & 2.189E-06 & 1.048E-06 & 1.722E-06 & 7.629E-07 \\
$^{64}$Cu+$^{64}$Cu& 3.835E-02 & 2.397E-02 & 4.181E-06  & 2.163E-06 & 3.731E-06 & 1.780E-06 & 2.783E-06 & 1.207E-06 \\
\hline \hline
\end{tabular} \label{biao2}
\end{table}

\begin{figure*}[!htbp]
\includegraphics[width=0.75\textwidth]{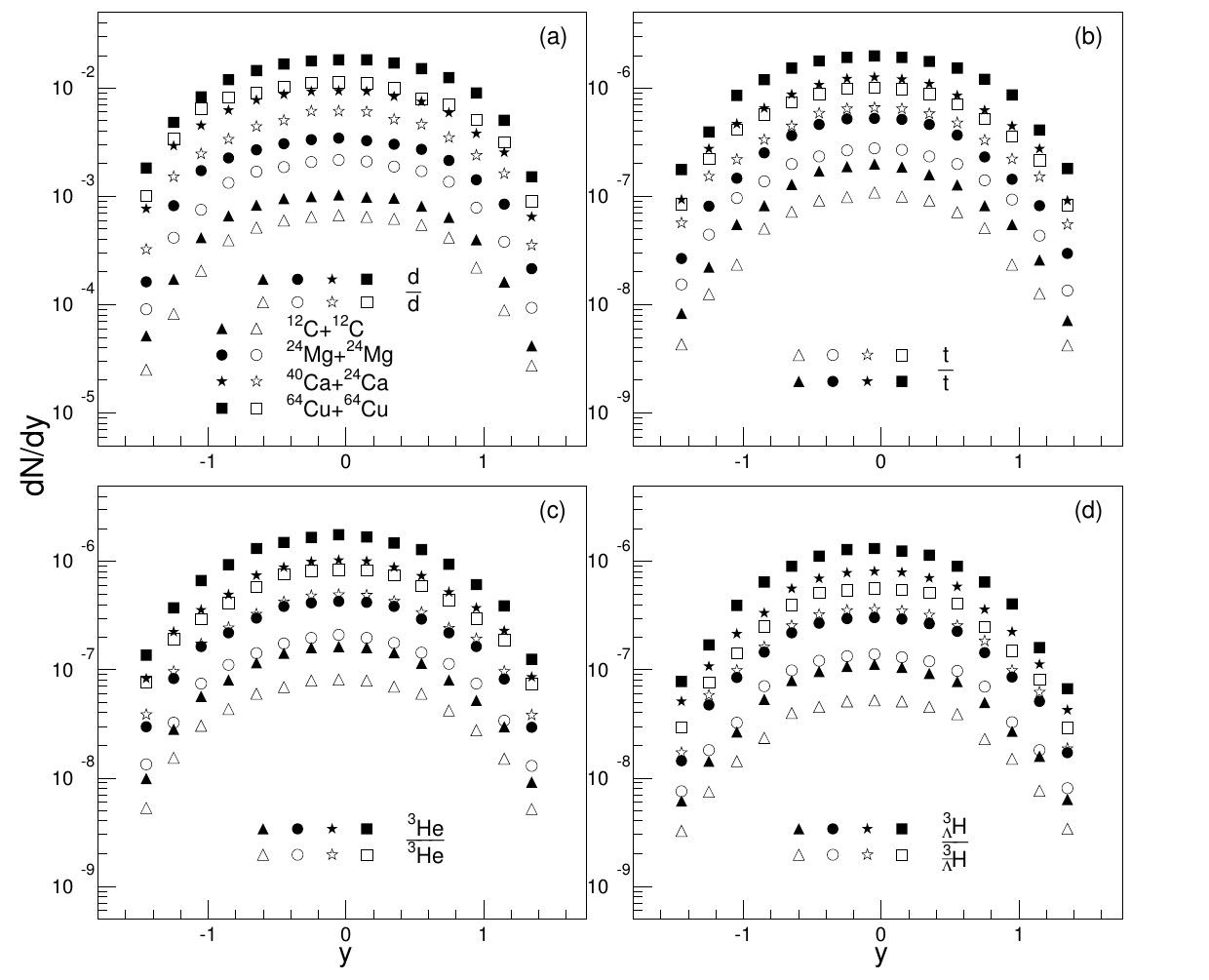}
\caption{The rapidity ($|y|$ $<$ 1.5) distributions of particles $d$ (a), $t$ (b), $\rm ^3He$ (c), and $\rm {_{\Lambda}^3H}$ (d) to corresponding antiparticles $\bar{d}$ (a), $\bar{t}$ (b), $\rm {\overline{^3He}}$ (c), and $\rm {\overline{_{\overline\Lambda}^3H}}$ (d) produced by PACIAE $+$ DCPC model at $\sqrt{S_{NN}}$ $=$ 200 GeV in $0-10\%$ centrality $^{12}$C+$^{12}$C, $^{24}$Mg+$^{24}$Mg, $^{40}$Ca+$^{40}$Ca, and $^{64}$Cu+$^{64}$Cu collisions.
}\label{tu2}
\end{figure*}

We show the rapidity distributions of $d$, $t$, $\rm ^3He$, $\rm {_{\Lambda}^3H}$ and their antinuclei in $^{12}$C+$^{12}$C, $^{24}$Mg+$^{24}$Mg, $^{40}$Ca+$^{40}$Ca, and $^{64}$Cu+$^{64}$Cu collisions at $\sqrt{S_{NN}}$ $=$ 200 GeV with PACIAE + DCPC models in FIG.~\ref{tu2}.
The particles and the antiparticles are mainly produced at the mid-rapidity and quickly decrease towards forward (backward) the rapidity. 
It is obviously that the rapidity distributions of antimatter are less than their corresponding matter with the same nucleus collisions. 
For the same particles produced, we find that the rapidity distributions raise with the nucleon number of the collision nucleus.

Then we compare the relationship between the nucleon number of the particles produced and their rapidity distributions in the same condition. 
We find that the rapidity distributions for $d$ are different from those for $t$, $\rm ^3He$, and $\rm {_{\Lambda}^3H}$, which show the less rapidity distributions than $d$ (a).
This is because that $d$ (a) contains two nuclei, while $t$ (b), $\rm ^3He$ (c), and $\rm {_{\Lambda}^3H}$ (d) all contain three nuclei. 
For the same nucleon number nuclei, $t$ (b) and $\rm ^3He$ (c), the former is consists of one proton and two neutron while the latter is consist of two proton and one neutron, we find that the rapidity distributions of $t$ (b) are roughly consistent with the $\rm ^3He$ (c).
However, the rapidity value of former is a little bit higher than latter. 
As shown in FIG.~\ref{tu2}, compared with the $t$ (b) (one proton and two neutron) and the $\rm {_{\Lambda}^3H}$ (d) (one proton, one neutron, and one $\Lambda$), we find that the rapidity distributions of the particles which has (anti-)hyperon is lower than that replaced by a neutron.

\begin{figure}[!htbp]
\includegraphics[width=0.75\textwidth]{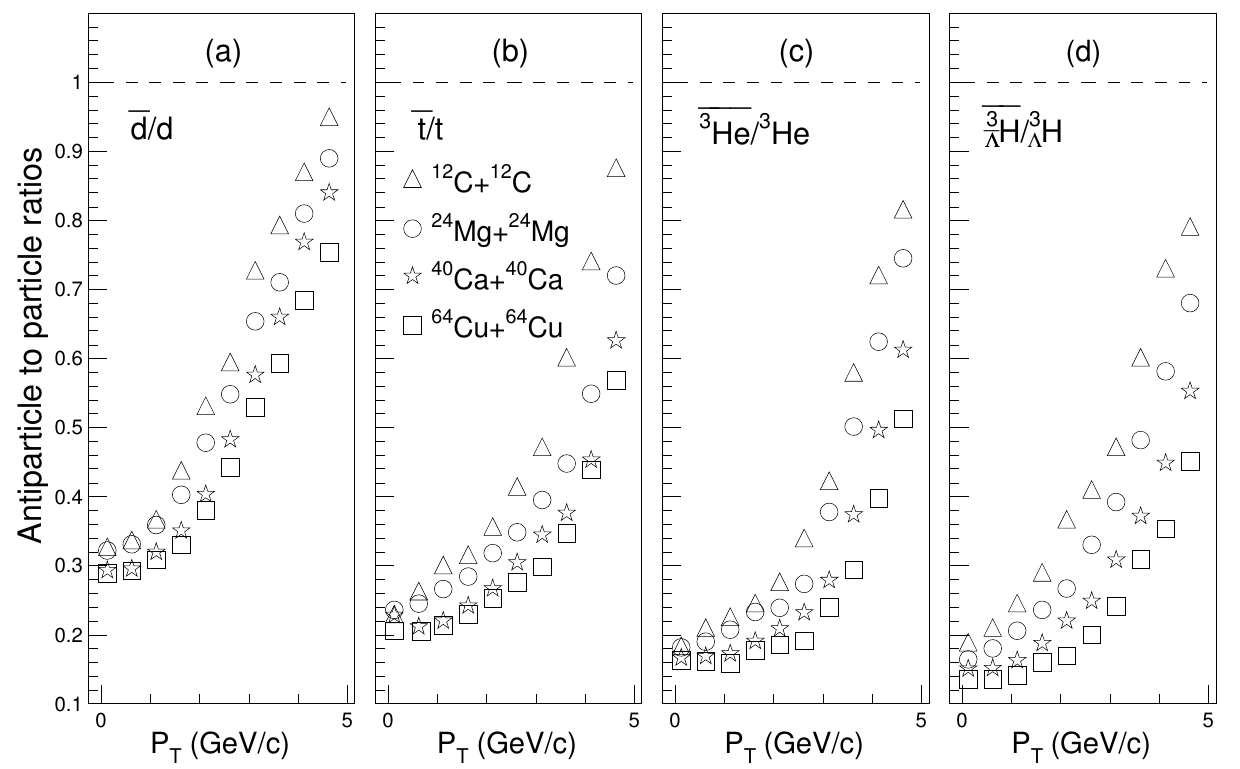}
\caption{The $p_{T}$ dependence of the ratios for antiparticles ($\bar{d}, \bar{t}, {\rm \overline{^3He}}$, and $\rm {\overline{_{\overline\Lambda}^3H}}$) to particles ($d$, $t$, $\rm ^3He$, and $\rm {_{\Lambda}^3H}$) produced by PACIAE $+$ DCPC model at $\sqrt{S_{NN}}$ $=$ 200 GeV in $0-10\%$ centrality $^{12}$C+$^{12}$C, $^{24}$Mg+$^{24}$Mg, $^{40}$Ca+$^{40}$Ca, and $^{64}$Cu+$^{64}$Cu collisions. 
The transverse momentum range is $0-5$ GeV/c.
}\label{tu3}
\end{figure}

The transverse momentum distributions of the ratios for $d$, $t$, $^3He$, and ${_{\Lambda}^3H}$ to their antiparticles produced in $^{12}$C+$^{12}$C, $^{24}$Mg+$^{24}$Mg, $^{40}$Ca+$^{40}$Ca, and $^{64}$Cu+$^{64}$Cu collisions are given in FIG.~\ref{tu3}. 
We find that in the same condition (the nucleon number and the particle production), the ratios of the antiparticle to particle are small at the low transverse momentum regions, while they have a rapidly increase at the high transverse
momentum regions. 
We also find that under the same transverse momentum, the ratios are inversely proportional to the nucleon number of the collision nucleus.

We also compare the relationship between the nucleon number of the particles produced and their antiparticle to particle ratios in the same condition. 
We find that the ratios of $\bar{d}/d$ (a) is higher than that of  $\bar{t}/t$ (b), $\rm {\overline{^3He}}/^3He$ (c), and $\rm {\overline{_{\overline\Lambda}^3H}}/{_{\Lambda}^3H}$ (d), because of the two nuclei $d$, and the others are three. 
For the same nucleon number nuclei, the antiparticle to particle ratios of  $\bar{t}/t$ (b), $\rm {\overline{^3He}}/^3He$ (c), and $\rm {\overline{_{\overline\Lambda}^3H}}/{_{\Lambda}^3H}$ (d) have a similar distributions.

\section {Conclusion}
\label{sum}
In this work, we use the PACIAE + DCPC models to investigate the productions of (anti)hypertriton and light (anti)nuclei generated by $0-10\%$ central $^{12}$C+$^{12}$C, $^{24}$Mg+$^{24}$Mg, $^{40}$Ca+$^{40}$Ca, and $^{64}$Cu+$^{64}$Cu collisions at $\sqrt{S_{NN}}$ $=$ 200 GeV with $|y|$ $<$ 1.5 and $p_{T}$ $<$ 5. 
We study the yield ratios of antiparticle to particle and the rapidity distributions of the different (anti)nuclei. The results obtained from our models are in agreement with the experimental data from the PHOBOS Collaboration.
It is obviously that the rapidity distributions of antimatter are less than their corresponding matter with the same nucleus collisions. 
The yield ratios of the particles in different transverse momentum regions are also given. 
We find that the ratios are raise with the increase of the transverse momentum, the ratios of the antiparticle to particle are small at the low transverse momentum regions while they have a rapidly increase at the high transverse momentum regions.

\begin{center} {ACKNOWLEDGMENT} 
\end{center}

This work is supported by the NSFC (Grants No.~11305144 and No.~11303023) and SPIE (Grant No.~201310491051) in China.

\bibliography{ref}

\end{document}